\documentclass[12pt]{article}
\usepackage{amsmath}
\usepackage{amssymb}
\usepackage{mathtext}
\usepackage{graphicx}

\begin{document}
\section*{DYNAMICS OF THREE VORTICES ON A PLANE AND A SPHERE --- III \\
Noncompact case. Problems of collaps and scattering}\footnote{REGULAR AND
CHAOTIC DYNAMICS V.3, No. 4, 1998\\
\it Received October 25, 1998\\
AMS MSC 76C05}

\begin{centering}
A.\,V.\,BORISOV\\
Faculty of Mechanics and Mathematics,\\
Department of Theoretical Mechanics
Moscow State University\\
Vorob'ievy gory, 119899 Moscow, Russia\\
E-mail: borisov@uni.udm.ru\\
V.\,G.\,LEBEDEV\\
Physical Faculty,\\
Department of Theoretical Physics
Udmurt State University \\
Universitetskaya, 1, Izhevsk, Russia, 426034 \\
E-mail: lvg@uni.udm.ru\\
\end{centering}

\begin{abstract}
In this article we considered the integrable problems of
three vortices on a plane and sphere for noncompact case.
We investigated explicitly the problems of a collapse and a scattering
of vortices and obtained the conditions of its realization.
We completed the bifurcation analysis and investigated
for collinear and Thomson's configurations the dependence of
stability in linear approximation and frequency of rotation in relative
coordinates from value of a full
moment. We indicated the geometric interpretation for characteristic situations.
We constructed a phase portrait and geometric projection for the integrable
configuration of four vortices on a plane.
\end{abstract}

\section{Motion of vortices on a plane}

Let us consider motion of vortices on a plane under the following condition
($a_k$ are inverse intensities of point vortices $a_k=1/\Gamma_k$)
\begin{equation}
\label{pl1}
A=A_1a_2 + a_2a_3 + a_1a_3\le 0\,.
\end{equation}
In case $A<0$ algebra of Poisson brackets is the algebra~$\mathbb{R} \oplus
so(2,1)$, and in case $A=0$ it is solvable algebra. As indicated in \cite{BL},
the symplectic sheet in both cases is noncompact (in the first
case the symplectic sheet is hyperboloid, in second it
is paraboloid), and  the trajectories of a representing
point on it can be both finite and infinite (this also
concerns the dynamics of three vortices in relative distances). In the
second case one can speak about a scattering of vortices.

Let us consider, at first, the case $A<0$. Without loss of
generality let us put $\Gamma_2, \Gamma_3>0$, $\Gamma_1<0$, $-\Gamma_1<\Gamma_2+\Gamma_3>0$.
Further, $D$ is the central
linear function of algebras of  Lie--Poisson
brackets ($D=\sum a_kM_k$, $M_k$
are quadrates of mutual distances between vortices).
From
\begin{equation}
\label{pl2}
\lambda ^2 = \frac{-3a_1a_2a_3}{D^2}(a_1+a_2+a_3)^3,
\end{equation}
for a stability coefficient of Thomson's configurations,
it follows, that the stability
of such configuration in a linear approximation
for various $D$ is determined
by sign of the sum of inverse intensities $S=\sum a_k$.
Therefore, we shall consider three separate cases corresponding to
the values $S>0$, $S<0$, and $S=0$.

The geometric interpretation and the bifurcation diagrams for various
characteristic combinations of parameters $A$, $S$ and $D$ are indicated
correspondingly in a Fig.~1 and Fig.~2.

{\it Case $A<0$ and $S>0$}.
The geometric interpretation shows that in this case for $D\le 0$ only finite
motions are possible for which trajectory on a plane of an integral of
the full moment is limited from both sides by the condition
$\Delta ^2>0$.

The bifurcation diagram (Fig.~2a) in this case reminds the diagram for vortices
on a plane in compact case for positive intensities with the exclusion,
that there is only one collinear configuration.
Another difference from
compact case is that the type of stability varies:
Thomson's configurations are stable for  positive
intensities in compact case and are unstable in this case (Fig.~3a),
and the collinear configurations are stable.
Both Thomson's, and the collinear solutions are determined only in area
$D>0$ (Fig.~1c). For positive values of the full moment both finite, and infinite motions are possible. By virtue of instability of Thomson's
solution, the solutions with any positive energy are allowed.

{\it Case $A<0$ and $S<0$}.
In this case for any value of the full moment the motion is
finite \mbox{(Fig.~1d--f)}.
Thomson's and the collinear solutions are determined
only for $D<0.$ That solutions are
remarkable because the energy of all
configurations tends to infinity with decrease of the
absolute value of the full moment down to zero (see Fig.~2b).
The curve of Thomson's configurations is situated above
of the curve for collinear configurations and
limits from above the area of possible movements. They
are accordingly stable and
unstable in a linear approximation (Fig.~3b).

{\it Case $A<0$ and $S=0$}.
For the value $D<0$ the tangency of the trajectory with the boundary of
area is possible. This trajectory corresponds to stable (see  Fig.~3c)
collinear configuration (Fig.~2c), which energy
tends to zero with decrease of $D$.
Thomson's solutions are presented only for $D=0$ and are degenerated, and have a
neutral equilibrium $(\lambda ^2=0).$ Such solutions are possible
for any distances between vortices and
fill the whole straight line  at Fig.~1h.
The dependence of an angular velocity from the distance between vortices is
determined by the formula \cite{Melesh}
\begin{equation}
\label{pl3}
\Omega = \frac{(\Gamma_1 + \Gamma_2 + \Gamma_3)}{2\pi M}\,,
\end{equation}
where $M$ is quadrate of the distance between vortices.
This straight line, corresponding to Thomson's
configurations, divides areas of scattering
and collapsing behavior of three vortices. Under a ``collapse'' we understand the
process of simultaneous collision of (in this case, three) vortices.
More detailed study of emergence of a collapse is indicated further.

For $D\ne 0$ the finite motions are presented for any value of the moment,
and infinite motions occurred only for negative values.

The dependence of an angular velocity of rotation both Thomson's and collinear
configurations in first  and second cases is qualitatively
identical and is defined by functions monotonically decreasing, with increase of
absolute value of a moment. For case $S=0$ the angular velocity of
rotation of a collinear configuration is slowly increased with
increase of the absolute value~$D$.

The last case corresponding to noncompact motion of vortices arises under
condition of $A=0$ (solvable algebra).
The motion is possible only for positive values of $D.$ There
are only two stationary configurations: one Thomson's and one collinear
(Fig.~2d). For  $D=0$ three vortices are placed on one straight line,
and each vortex is placed at center of a vorticity  of two remaining. It is a corollary of
general possibility of reducing of a problem of $n+1$
vortices to problem of $n$ vortices.

\section{Motion of vortices on a sphere}

Similarly to Section 1 we shall construct the bifurcation diagrams
for the sphere, and also plots of absolute motion, using for small $D$
appropriate dependencies for a plane and the method of a continuation on the
parameter. The values $d_k=4R^2(a_i + a_j)$ determine those
values of the full moment, which can arise in problem of two vortices. The
geometric interpretation for motion on the sphere is indicated in a Fig.~5.

{ \it Case $A<0$ and $S>0$}. In addition to Thomson's and
collinear configurations
that exists in plane case, for  $D>0$ there are two collinear
configurations, appearing from a problem of two vortices
(Fig.~4a). By continuation on the moment one of them merges with
Thomson's,   which   disappears   for   value~$d_T$.    The    collinear
configuration continues and merges  with another collinear
configuration, that  has  an  analog  on  the  plane.  After  this  both
configurations disappears also.
For $D>d_3$ and $D<d_1$ the stationary configurations do not exist.
Thomson's configuration are in this case always unstable (Fig.~6a), and the
collinear configurations are stable.

{ \it Case $A<0$ and $S<0$}. In this case the confluence of Thomson's and
one of collinear (closest to Thomson's on energy) configurations (Fig.~4b)
are happened. Besides there are collinear configurations from a problem of
two vortices (for one of them $E\mapsto \infty $ by coincidence of
vortices because of existence of one negative intensity). One of them
extended into area of positive values of $D$. Thomson's configuration, as
well as in plane case, is stable in the linear approximation (Fig.~6b).
Some of collinear configurations are also stable. One of them becomes
stable in confluence with Thomson's configuration, and other loses
stability in confluence with unstable collinear configuration.

{\it Case $A<0$ and $S=0$}.
Instead of one collinear configuration in case of a plane
in this case there are three various branches
of collinear configurations, as for $D>0$ and for $D<0$ (see Fig.~4c).
The solution with higher energy is unstable for $D>0$
and disappears merging with collinear configuration with lower energy.
The solution with the lowest energy is stable.
For $D<0$ situation is reversed.
Thus collinear solution with lower energy merges with the solution,
appeared at $D=d_2.$
As well as on the plane, the Tomson solution are present only for $D=0$.
In this case all trajectories are collapsing (Fig.~5h).

At last, under condition of $A=0$, the bifurcation diagram is indicated in a
Fig.~4d. As well as in plane case, the stationary solutions exist only for
$D>0$, (Fig.~5 j--l). The difference consists in emerging of collinear
configurations of a problem of two vortices.
One such configuration merges with Thomson's, then both disappear.
All collinear solutions are stable, and Thomson's is unstable (Fig.~6d).

The dependence of angular velocity of rotation for the listed situations is
represented in a Fig.~7. As well as in compact case, the frequency of
rotation increasing with creation of the third vortex from the problem of two
vortices.

{\bf Remark 1.} {\it Such magnification, as well as in compact case, is
explained by the fact, that for the birth of a pair of vortices from one
vortex of total intensity there is a rotation of a collinear configuration
around an axis, which is passing through the third vortex. In the result
this vortex actually does not influence on rotation of two appearing
vortices, which angular velocity depends on the mutual distance between
them as}
\begin{equation}
\Omega = \frac{1}{2\pi M_3}
\sqrt{(\Gamma_1 + \Gamma_2)^2-\frac{\Gamma_1\Gamma_2 M_3}{R^2}}\,,
\end{equation}
{\it Where $M_3$ is quadrate of the distance between first two vortices, and
$\Gamma_1, \Gamma_2$ are intensities, correspon\-ding to it. As the last
expression shows the angular velocity tends to infinity with $M\mapsto
0$.}

The study of dynamics of angles of declination of rotation axis to the plane of
vortices shows, that in first and second listed
cases the angle of declination of axes increasing
from zero up to $\pi/2$. So, therefore the
rotation axis tends to take a position in the plane of vortices and appropriate
Thomson's a configuration turned into collinear (Fig.~8).

{\bf Remark 2.} {\it
Note, that a condition of existence of static configurations on a
sphere for a specific collection of intensities is}
\begin{equation}
\label{sp1}
a_1a_2a_3 (a_1 + a_2 + a_3)>0\,,
\end{equation}
{\it which determines stability of Thomson's configuration}
\begin{equation}
\label{sp2}
\lambda ^2 = \frac{D-3R ^ 2 (a_1 + a_2 + a_3)}{9D^2}a_1a_2a_3 (a_1 + a_2 +
a_3)\,.
\end{equation}
{\it This correlation is easy for understanding from the geometric interpretation
(see Fig.~5d), which illustrates a possibility of reaching of function $H$
to its extremum. Let us note, that as well as
in compact case, that was considered in \cite{BL}, the static
configurations on a sphere are observed only under
fulfillment of the condition $\Gamma_1\Gamma_2 + \Gamma_2\Gamma_3 +
\Gamma_1\Gamma_3>0$, (Fig.~7).}

\section{The condition of a collapse of vortices on a plane and a sphere}

Let us consider conditions of emergence of a collapse of
three vortices on a plane
and a sphere. It is known, that the collapse is impossible for
any pairs of vortices taken separately, because with the approach of
such pair influence of
the third (remote) vortex negligible small, and two vortices on a
plane (sphere) move relatively each other so,
that the distance between them did not change. The necessary condition of
the collapse is the fulfillment of $D=0$, since it follows $M_k=0$.
The condition $D=0$ allows to proceed from four-dimensional to
three-dimensional Lie algebra, for which actual dynamics
takes place on the singular
symplectic sheet, so that the nontrivial singular symplectic
sheet (surface of a
cone) will correspond only to algebra
$so(2,1)$. The similar statements are valid also for a simultaneous collapse $n$
vortices, which is not almost investigated (except for cases $n=4,5$)
\cite{Nov, Melesh}.

The conditions of a collapse of vortices on a sphere, obviously, will coincide
with conditions for a plane, since on small distances the influence
of curvature to dynamics of vortices is insignificant.

{\bf Remark 3.}{\it
The problem of the collapse is one of the most
interesting problems connected to
vortices and represents large interest for the theoretical hydromechanics
as one of the
models, which can be used for understanding of the transition to the
turbulence contained in a nonuniqueness of
solutions of the hydrodynamic equations of the Euler. Really,
the theorem of existence and uniquenesses for these equations are proved in the
supposition of sufficient smoothness of an initial field of velocities. From
a mathematical point of view the process of the collapse of
vortices representing
confluence of the special solutions of Euler equation of $\delta$-function type, under inverting of time will determine disintegration of
vortices with corresponding loss of uniqueness.
Therefore large interest represents the study of this
problem from the point of view of a
regularization of collisions in the similar way, how it
is done in classical celestial
mechanics \cite{ArnoldKozlovNei}.
For physics of atmosphere the phenomenon of the collapse can be considered as a
model of forming of large atmospheric vortices.}

We research at first possibility of
a homogeneous collapse for three vortices.
For the homogeneous collapse all distances between vortices depend on time in the identical way
and are in the constant proportion \cite{Melesh}.
Using isomorphism with a problem of Lottka--Voltera
(both for a plane, and a sphere~\cite{BP}), we
shall consider a homogeneous system of equations in the  form
\begin{equation}
\label{c1}
\hookrightarrow \dot M_1 = \Gamma_1 M_1(M_2-M_3)\,.
\end{equation}
The homogeneous asymptotic solutions (\ref{c1}) we shall search as
\begin{equation}
\label{c2}
M_k =\frac{C_k}{\tau}\,.
\end{equation}
Let's note, that such solutions are used in Kovalevskaya method for
construction of full-parametric Laurent expansion.
The nontrivial solutions of such form are possible only under condition of
$S=0$.
If this condition is fullfiled, the solution can be written as
\begin{equation}
\label{c3}
M_1=\frac{C}{\tau}\,, \qquad M_2=\frac{C+a_3}{\tau}\,,\qquad M_3=\frac{C-a_2}{\tau}\,,
\end{equation}
where $C$ is arbirary constant.

The solution (\ref{c3}) is full-parametric and contains only collapsing and
running up (for a plane) trajectories. However, there is still the special solution
$M_1=M_2=M_3=const$, describing the Thomson's configuration, which in this
case are degenerate. Using the relation
\begin{equation}
\label{c4}
dt = \frac{2\pi M_1M_2M_3}{ \Delta}d\tau\,,
\end{equation}
it is possible to receive an asymptotics of the
solution (\ref{c3}) in real time
\begin{enumerate}
\item for a plane: $t=\frac{D}{\tau}$, $M_k=C_k^*t;$
\item for a sphere: $t=AR^2\left(1-\sqrt {1-\frac{B}{
\tau R^2}}\right)$, $M_k=D_kR^2\left(1-\left(1-\frac{t}{AR
^2}\right)^2\right)$,
\end{enumerate}
where $A, B, C_k^*,D, D_k=\text{const}$, $R$ is radius of a sphere.
For a sphere the absolute motion of vortices for conditions $D=0$ and $S=0$
consists in scattering of vortices from one point up to the
moment of reaching of equator and further approach in some other point.
Let us analyze the possibility of the collapse in the
system of three vortices on a plane in general inhomogeneous
case. Writing the necessary condition of the collapse
$D=0$ in absolute variables
\begin{equation}
\label{c5}
\left(\sum \Gamma_k\right)I-P^2-Q^2=0\,,
\end{equation}
where the uninvolute integrals of motion $P,Q,I$ are equal
\begin{equation}
\label{c6}
P = \sum \Gamma _kx_k\,,\qquad Q = \sum \Gamma _ky_k\,,\qquad
I = \sum \Gamma _ k(x_k^2 + y_k^2)\,,
\end{equation}
we found two special cases $ \sum \Gamma _ k = 0 $
and $\sum \Gamma _k \ne 0$. Without loss of generality
let us assume $\Gamma _1=-1$, $ \Gamma _2,\Gamma _3>0$.
Under conditions $ \sum \Gamma _k = 0$, from
(\ref{c6}) it follows that $P=Q=0$. It means, that the third
vortex is at center of vorticity of first two vortices, rotating
uniformly with frequency
$$
\Omega = \frac {1}{2\pi} \frac{\Gamma _1 +\Gamma _2}{M_{3}}\left(1+\frac{\Gamma
_1}{\Gamma _2}+\frac{\Gamma _2} {\Gamma _1}\right)\,,
$$
around the point with the  radius-vector
$$
\vec r = \frac{\Gamma '_1 {\vec r_1}+\Gamma '_2{\vec r_2}}{\Gamma '_1 + \Gamma
'_2}\,, \qquad
\Gamma '_1 = \Gamma _1-\frac{(\Gamma _1+\Gamma _2)^2}{\Gamma _1}\,, \qquad
\Gamma '_2 = \Gamma _2-\frac{(\Gamma _1 + \Gamma _2)^2}{\Gamma _2}\,,
$$
where $M_{3}$ is quadrate of the distance between the first
and second vortices, and
$\vec   r_k,\,\Gamma _k$   are   the   radius-vector   and   intensity
correspondingly.
Therefore, collapse for case of zero total intensity is impossible.

Under condition of $ \sum \Gamma _k \ne 0$ the collapse is possible only under
condition of noncompactness of algebra of vortices $A=a_1a_2-a_1-a_2<0$, as in
compact case a symplectic sheet (the sphere) does not pass through a
beginning of coordinates $M_k=0$.

For the determination of sufficient conditions of the collapse we shall consider the
projection of trajectories on a plane $M_1,M_2$. Let us express $M_3$ from
the integral of the full moment (for $D=0$)
\begin{equation}
\label{c7}
M_3=a_1M_1+a_2M_2\,.
\end{equation}
The physical area on a plane $M_1,M_2$ for $D=0$ is limited by straight lines, passing
through the zero with coefficient of declination
\begin{equation}
\label{c8}
K_{1,2}=\left (\frac{1\pm \sqrt {-A}}{1-a_2} \right)\,.
\end{equation}
The trajectory of regularized system is determined by the relation (\ref{c7}) and
equation for energy of the system, which can be presented as
\begin{equation}
\label{c8}
M_1^{-a_1}M_2^{-a_2}(a_1M_1+a_2M_2) = \text{const}(E)\,.
\end{equation}
The analysis of the equation (\ref{c8}) shows, that for various values
of parameters $a_1, a_2$ there are three various kinds of trajectories:
\begin{enumerate}
\item if $a_1+a_2<1$, then all trajectories are compact,
looks like closed loops, which are going out the origin
of coordinates, tangentcing axes $OM_1$ and
$OM_2$ (Fig.~9a, curves b, c);
\item if $a_1+a_2>1$, then all trajectories are noncompact,
going to infinity, asymptotically press oneself to axeses $OM_1$ and $OM_2$;
\item if $a_1+a_2=1$, then all trajectories are straight lines
which are going out from the origin of coordinates under
different angles (the Fig.~9a, curve a);
\end{enumerate}

Let us mark on a plane of parameters $a_1,a_2$, areas appropriate to given
types of trajectories. It is necessary to mark on this plane also values
of parameters describing various forms of physical area.
By (\ref{c8}), the physical area for $a_1\ne 1,a_2\ne 1$ (such, that
$A<0$), represents an interior of an sharp angle located inside a
quadrant $M_1>0,M_2>0$. If $a_1=1$ ($a_2=1$), one of the sides of physical
area coincides with an axes $OM_2$ ($OM_1$). In case $a_1=a_2=1$ the motion is
allowed in the whole quadrant $M_1>0$, $M_2>0$.

Comparing possible types of trajectories (for $A<0$, $D=0$) with types of areas
of motion with consideration, that on reaching of the boundary motion continues on the same
trajectory in the opposite direction, we conclude:
\begin{enumerate}
\item In the system of three vortices the homogeneous collapse
(scattering) is possible, it happens when the relation $a_1+a_2=1$ is
executed for inverse intensities;
\item Scattering of vortices is possible only under conditions $a_1=1$, $a_2=1$,
$a_1=a_2=1$ (in this case the vortex motion never passes through the
collinear configuration);
\item For other values of inverse intensities $a_i$ the motion
of vortices is bounded between two collinear configurations, and the distance between them is
limited (see Fig.~9b).
\end{enumerate}

\section{Scattering of vortices on a plane}

In the system of three vortices we shall name scattering trajectories, such trajectories that
at least one of mutual distances $(M_1, M_2, M_3)$ is infinitely
increased. In difference from the collapse, the scattering can happen for
$D\ne 0$.

Let us define new variables, with the help of which scattering problem of
vortices on a plane can be reduced to the research of collapse:
\begin{equation}
\label{rs1}
\hookrightarrow
X_1 = \frac{1} {M_2M_3}\,.
\end{equation}
Their equations of motion have a form
\begin{equation}
\label{rs2}
\hookrightarrow
\dot x_1 =x_1 ((\Gamma _2-\Gamma _3) x_1+\Gamma _3x_2-\Gamma _2x_3)\,,
\end{equation}
and the bounding physical area relation $\Delta ^2\ge 0$, will not be changed
\begin{equation}
\label{rs3}
2(x_1x_2 + x_1x_3 + x_2x_3) -x_1^2-x_2^2-x_3^2\ge 0\,.
\end{equation}
Let us select, as above $a_3=-1$, $a_1,a_2>0$. The trajectory in
space of variables $x_1,x_2,x_3$ is set by integrals of the
moment $D$ and the energy $E$, which can be presented in the form
\begin{equation}
\label{rs4}
{ a_1x_1 + a_2x_2 + a_3x_3 = D\sqrt {x_1x_2x_3}\,,}
\end{equation}
\begin{equation}
{x_1^{a_2-a_1-1}x_2^{a_1-a_2-1}x_1^{1+a_1+a_2}=C(E)\,,}
\end{equation}
where $C(E)$ is some function of energy.

For $D=0$ the type of trajectories, defined by (\ref{rs4}), is similar to the one investigated above. In this case to scattering 2) in a system of three vortices corresponds an inhomogeneous collapse in a system (\ref{rs2}), for  conditions
$a_1=1$, $a_2=1$, $a_1=a_2=1$ (see Fig.~9b).

It is interesting to note, that the homogeneous collapse and the
scattering in  variables $M_k$ remains homogeneous in
variables  $x_k$, only direction of motion on trajectories varies.

Obvious quadratures and the analysis of absolute motion is in this case
indicated in \cite{Melesh}.

The numerical research for $D\ne 0$ shows, that in variables  $M_k$ there are
only trajectories of types 1), 3) of the previous section.
In this connection, it seems, that
the availability of vortex pairs ($a_1=1$, $a_2=1$, $a_1=a_2=1$) in the
system is necessary and sufficient condition of scattering.

{\bf Remark 4.} {\it
In a work \cite{Eckh} reducing of the order is executed and
the phase portraits of integrable case of a problem with zero
total intensity and zero total moment are indicated. In this work
the canonical form of equations of motion is used. For
the indicated conditions on a level of algebra of brackets there
is an reduction of a problem of four vortices to a problem of three
vortices with the reduced Hamiltonian. On a
symplectic sheet of three-dimensional vortex algebra, defined by
introduced (as a corollary of the integrability) an invariant relation
such as an integral of the moment with a constant~$D_1$, standard
canonical variables $L,l$ are presented. The condition of compactness of a symplectic sheet
will have a very simple form, if three vortices of four have intensities
of same sign. In compact case, for various values of intensities the
phase portraits are represented at Fig.~10. In a general situation of unequal
intensities there are six collinear stable solutions and three uncollinear
unstable ones. The last solutions generalize of Thomson's  configuration, however
distances between vortices are not equal. The connection between
energy $E$ and  moment  $D_1$  for  these  solutions  is  determined  by
relation}
$$
E=F(\Gamma _1, \Gamma _2, \Gamma _3) D_1^{\frac {1}{8\pi} (\Gamma _1^2 + \Gamma
_2^2 + \Gamma _3^2)}\,,
$$
{\it where $f(\Gamma _1, \Gamma _2, \Gamma _3) $ is some function dependent on
intensities.
The bifurcation analysis, consisting in a determination
of an explicit  form of
function $f(\Gamma _1, \Gamma _2, \Gamma _3)$, can be executed similarly to
Section~1.
In case if there are only two intensities of the same sign, the
symplectic sheet is noncompact and the scattering is possible.
The regular scattering, for example, is possible in case of
interaction two (generally speaking, various)
vortex pairs \cite{Eckh1}. As far as we know, in general
case the conditions of a collapse and a scattering in considered problem are not
investigated.}

After submission of the article in the journal the authors have received
prints of  articles  \cite{Mars,New1,New2}  from  J.E.Marsden  and  P.K.Newton.
Their results were  received  with the help of the ``canonical'' approach
and are contained some results of our work. We are  extremely  grateful  to
them. The authors thank I.\,S.\,Mamaev and N.\,N.\,Simakov for
useful discussions and help in
work. The work is carried out under the support of
Russian Fond of Fundamental
Research (96--01--00747) and Federal program ``States
Support of Integration
High Education and Fundamental Science'' (pro\-ject~No.~294).

\begin{figure}[ht!]
\begin{center}
\begin{tabular}{ccc}
\includegraphics{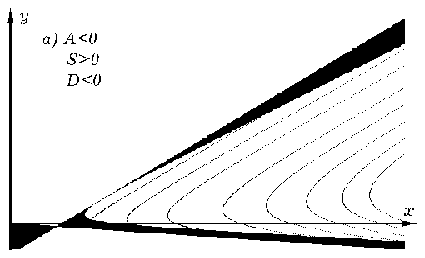} & \includegraphics{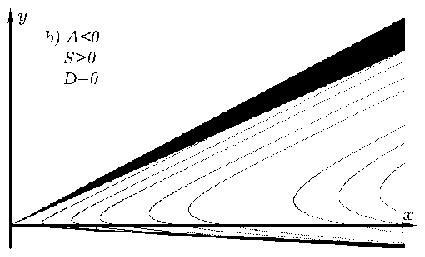} &
\includegraphics{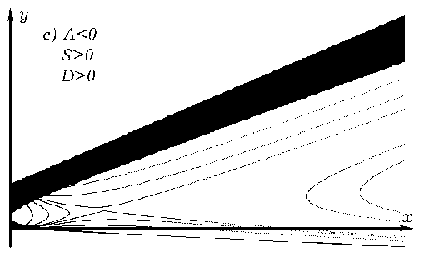}\\

\includegraphics{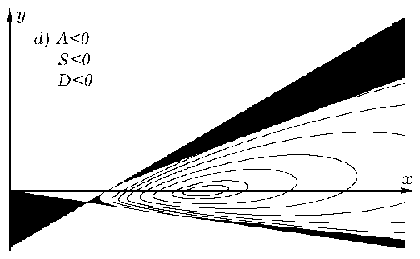} & \includegraphics{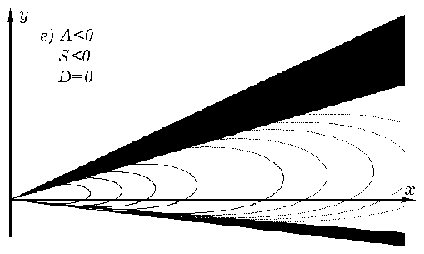} &
\includegraphics{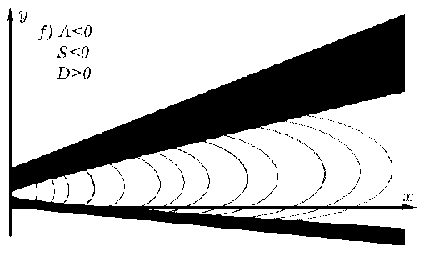}\\

\includegraphics{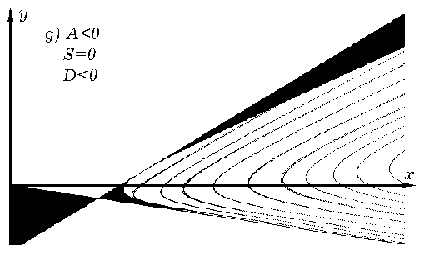} & \includegraphics{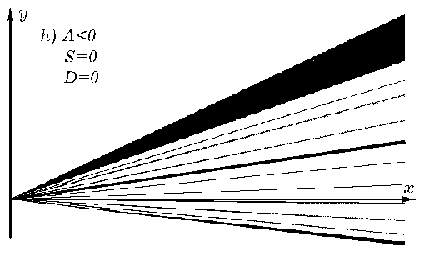} &
\includegraphics{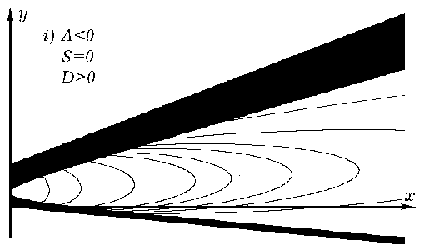}\\

\includegraphics{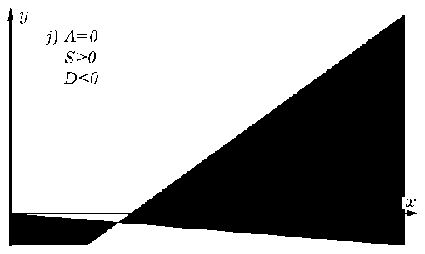} & \includegraphics{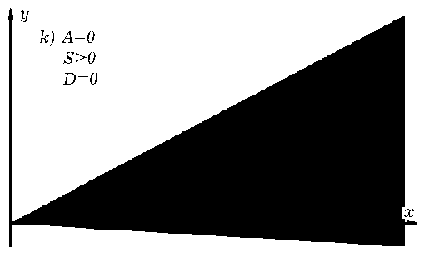} &
\includegraphics{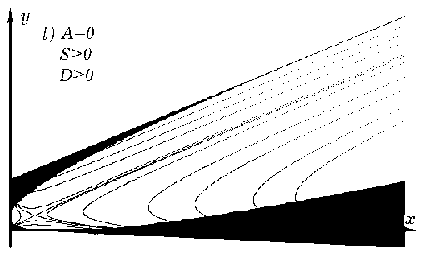}\\
\end{tabular}
\end{center}
\caption{The geometric interpretation for various values of parameters $A$, $S$,
$D$. The dark color designates area of positive values $M_k\ge 0$,
for which $\Delta ^2>0$. For figure h) the bold line corresponds to a set of
Thomson's configurations.}
\end{figure}

\begin{figure}[ht!]
\begin{center}
\begin{tabular}{cc}
\includegraphics{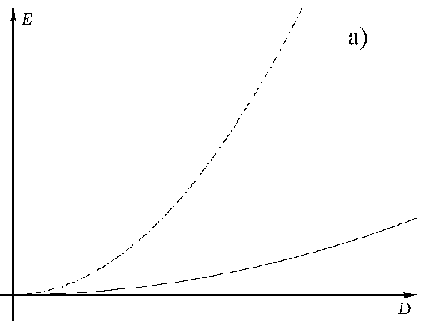} & \includegraphics{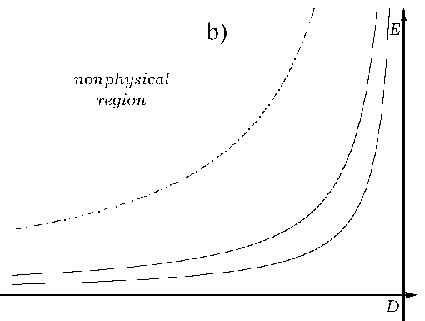}\\
\includegraphics{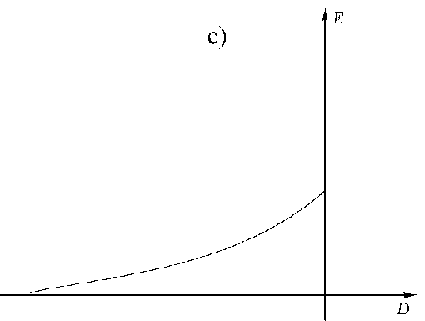} & \includegraphics{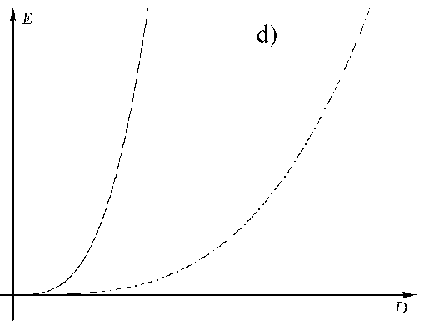}\\
\end{tabular}
\end{center}
\caption{Bifurcation curves on a plane for cases
a) $A<0$ and $S>0$;
b) $A<0$ and $S<0$;
c) $A<0$ and $S=0$;
d) $A=0$ and $S>0$. Here and further dot-dashed line corresponds to Thomson's
configurations, and solid lines to collinear.}
\end{figure}

\begin{figure}[ht!]
\begin{center}
\begin{tabular}{cc}
\includegraphics{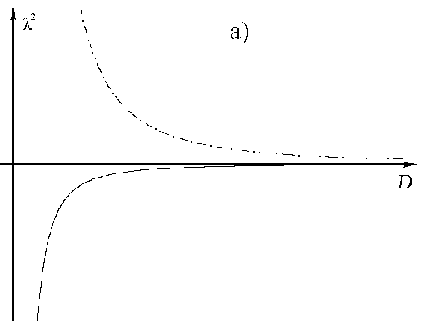} & \includegraphics{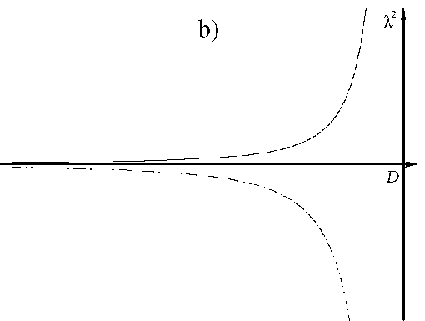}\\
\includegraphics{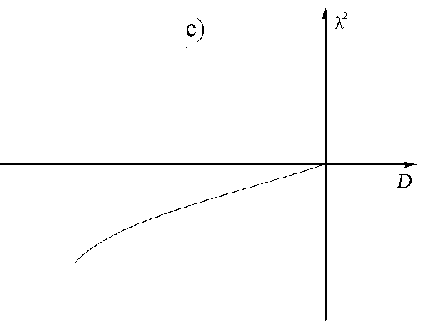} & \includegraphics{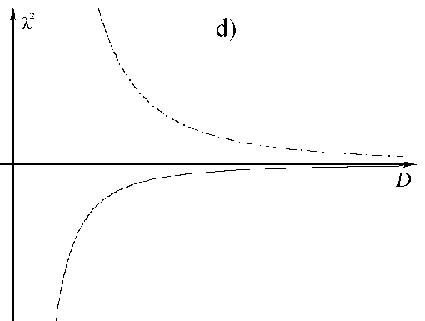}\\
\end{tabular}
\end{center}
\caption{The stability coefficient on a plane for cases
a) $A<0$ and $S>0$;
b) $A<0$ and $S<0$;
c) $A<0$ and $S=0$;
d) $A=0$ and $S>0$.}
\end{figure}

\begin{figure}[ht!]
\begin{center}
\begin{tabular}{cc}
\includegraphics{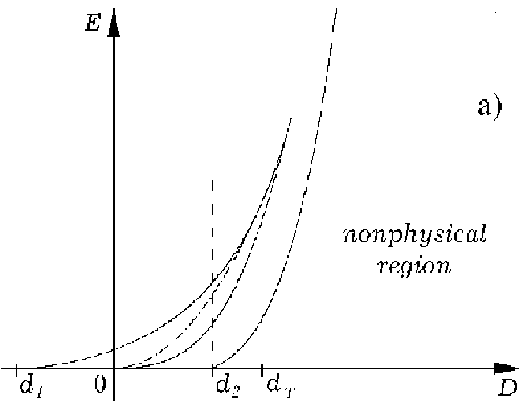} & \includegraphics{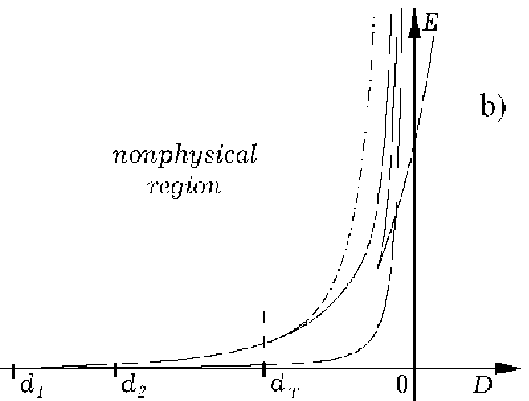}\\
\includegraphics{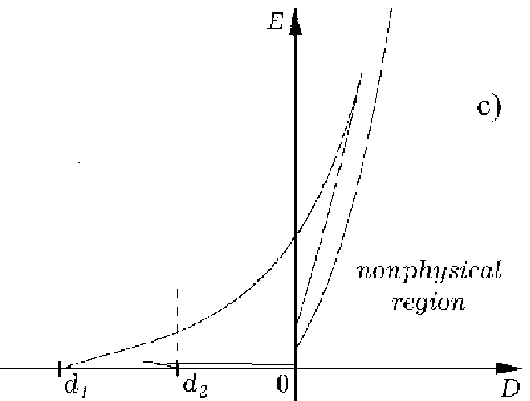} & \includegraphics{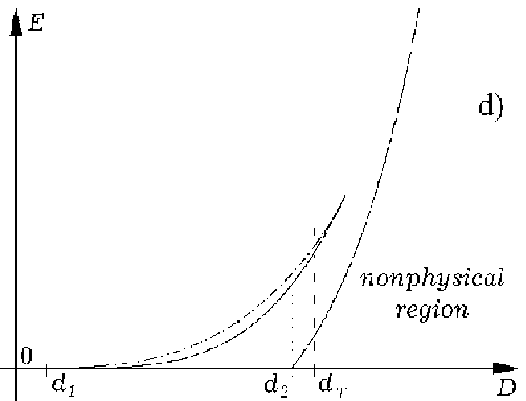}\\
\end{tabular}
\end{center}
\caption{Bifurcation curves on a sphere for cases
a) $A<0$ and $S>0$;
b) $A<0$ and $S<0$;
c) $A<0$ and $S=0$;
b) $A=0$ and $S>0$.}
\end{figure}

\begin{figure}[ht!]
\begin{center}
\begin{tabular}{ccc}

\includegraphics{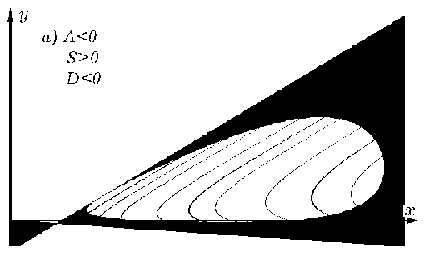} & \includegraphics{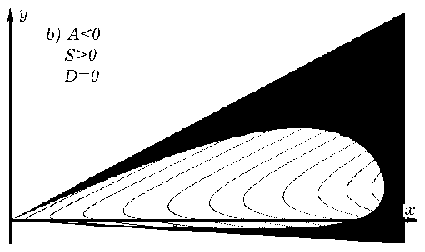} &
\includegraphics{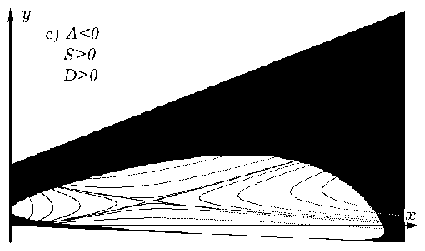}\\

\includegraphics{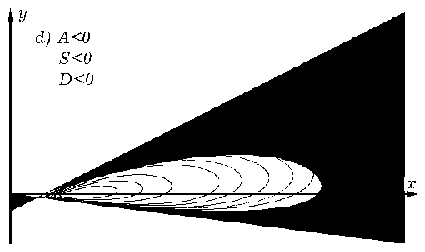} & \includegraphics{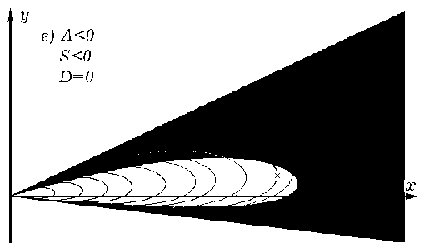} &
\includegraphics{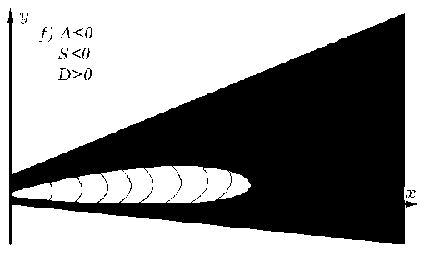}\\

\includegraphics{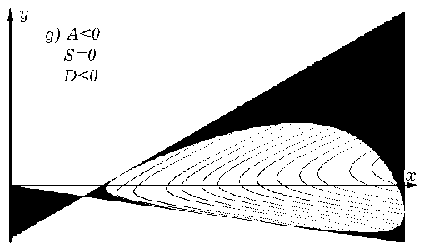} & \includegraphics{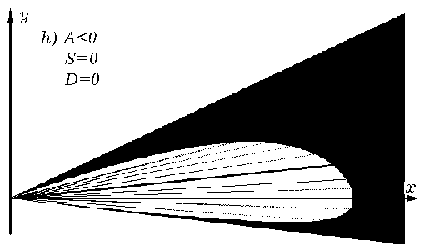} &
\includegraphics{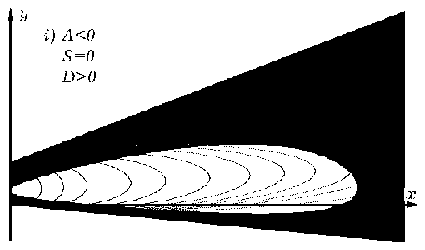}\\

\includegraphics{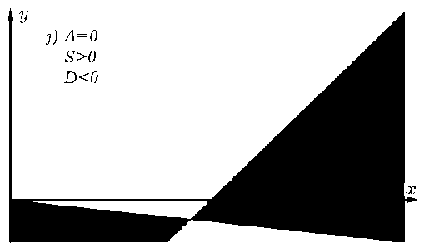} & \includegraphics{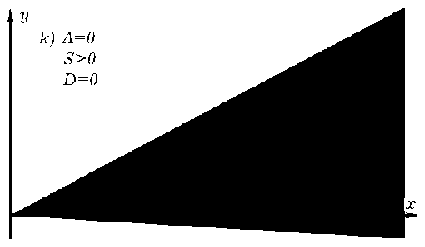} &
\includegraphics{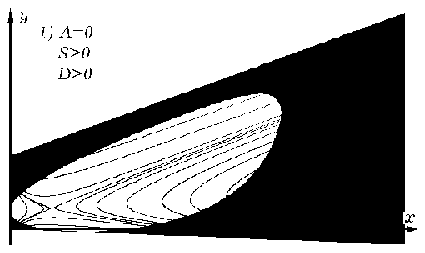}\\
\end{tabular}
\end{center}
\caption{The geometric interpretation for a sphere for various values of
parameters $A$, $S$, $D$. The dark color designates area of positive
values $M_k\ge 0$, which for $\Delta ^2>0$.}
\end{figure}

\begin{figure}[ht!]
\begin{center}
\begin{tabular}{cc}
\includegraphics{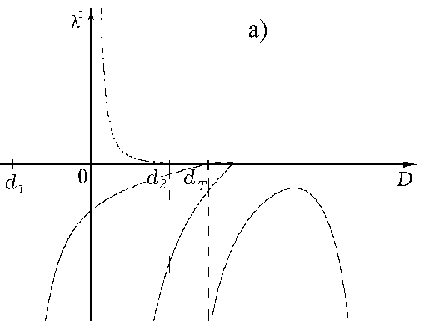} & \includegraphics{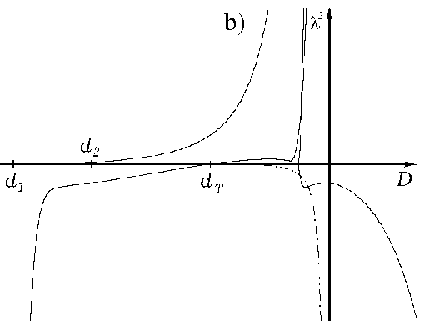}\\
\includegraphics{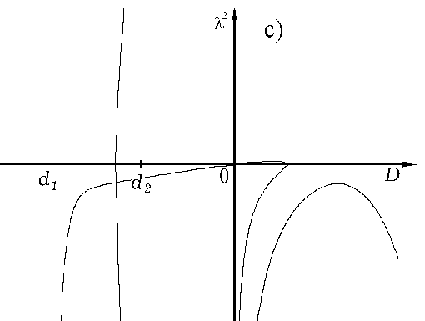} & \includegraphics{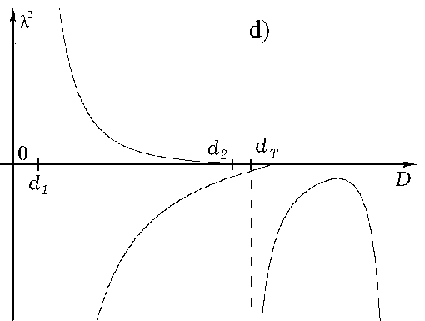}\\
\end{tabular}
\end{center}
\caption{Stability coefficient on a sphere for cases
a) $A<0$ and $S>0$;
b) $A<0$ and $S<0$;
c) $A<0$ and $S=0$;
b) $A=0$ and $S>0$.}
\end{figure}

\begin{figure}[ht!]
\begin{center}
\begin{tabular}{cc}
\includegraphics{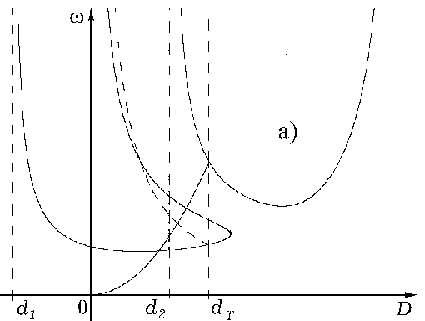} & \includegraphics{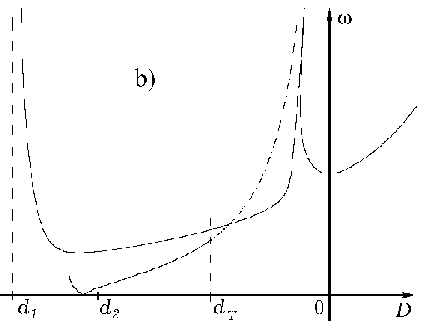}\\
\includegraphics{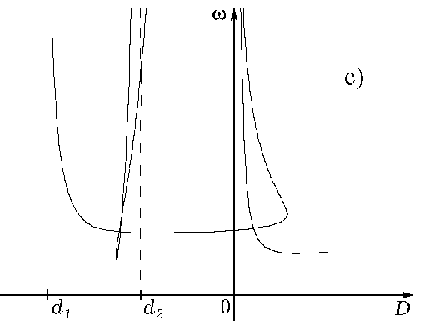} & \includegraphics{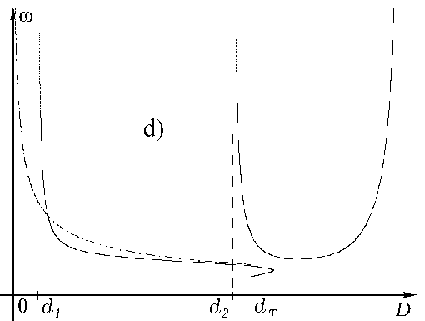}\\
\end{tabular}
\end{center}
\caption{An angular velocity on a sphere for cases
a) $A<0$ and $S>0$;
b) $A<0$ and $S<0$;
c) $A<0$ and $S=0$;
b) $A=0$ and $S>0$.}
\end{figure}

\begin{figure}[ht!]
$$
\includegraphics{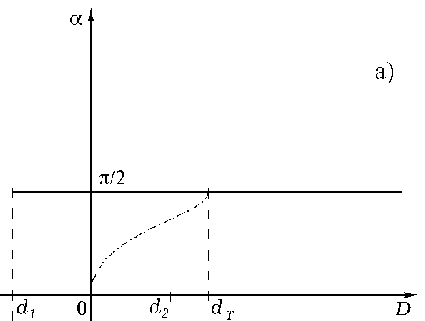}
\includegraphics{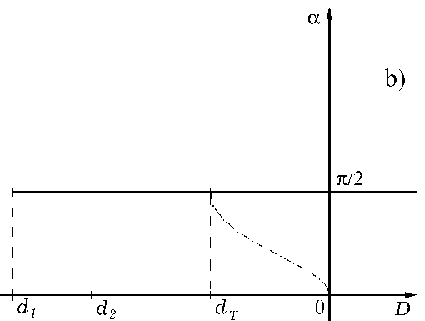}
$$
\caption{The angle of declination on a sphere for cases
a) $A<0$ and $S>0$;
b) $A<0$ and $S<0$.}
\end{figure}

\begin{figure}[ht!]
$$
\includegraphics{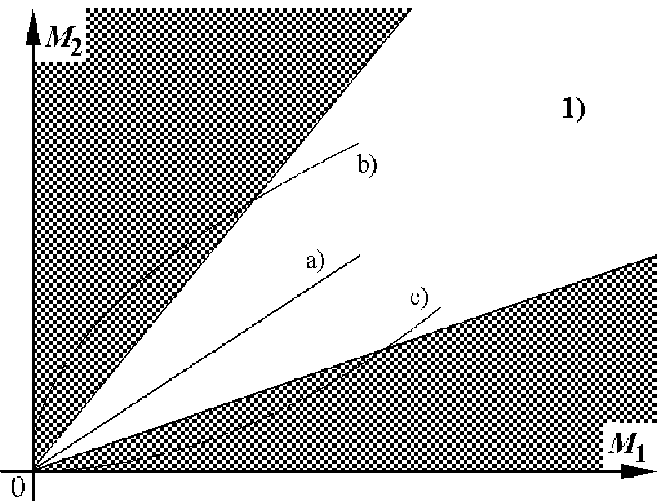}
\includegraphics{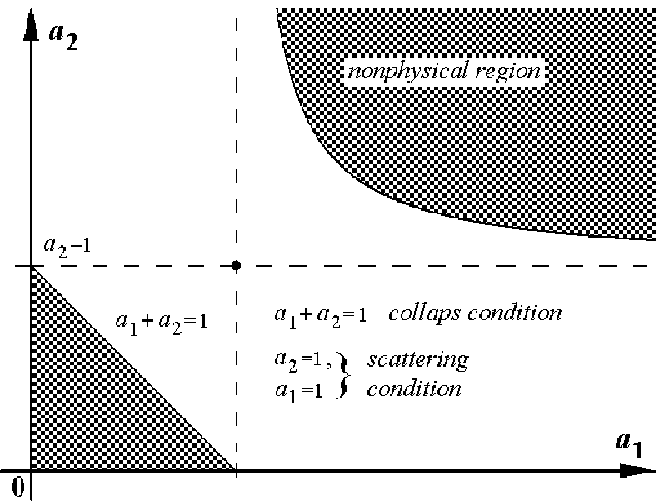}
$$
\caption{1) Possible asymptoticses of a behavior of trajectories of
the system of three vortices near to zero on a plane $(M_1, M_2)$
in case of the collapse (the nonphysical area is shaded);
2) Planes of parameters ($a_1, a_2$
for $a_3=1$).
The collapse is possible on straight line $a_2=a_1-1$,
scattering corresponds to asymptotes $a_1=1, a_2=1$.}
\end{figure}

\begin{figure}[ht!]
$$
\includegraphics{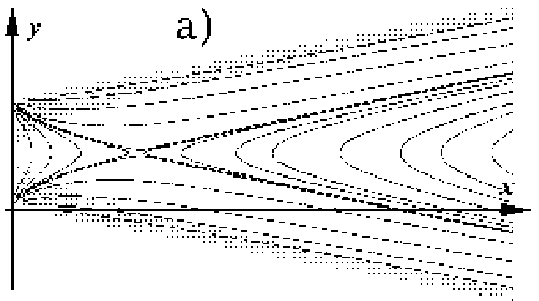}
\includegraphics{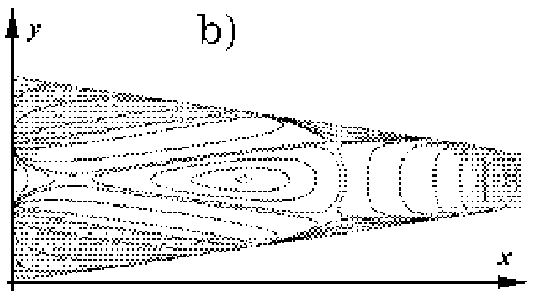}
\includegraphics{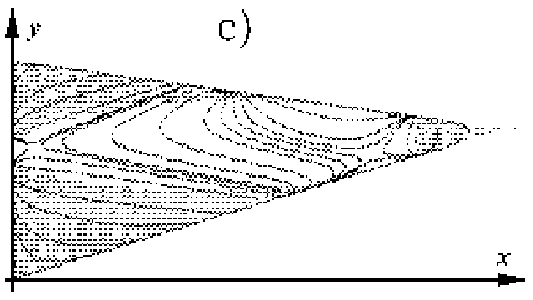}
$$
\caption{The geometric projection for case of four vortices on a plane
with zero full intensity for
a) $a_1=-a_2=a_3=-a_4$;
b) $a_1=a_2=a_3\ne a_4$;
c) $a_1\ne a_2\ne a_3$.}
\end{figure}

\begin{figure}[ht!]
$$
\includegraphics{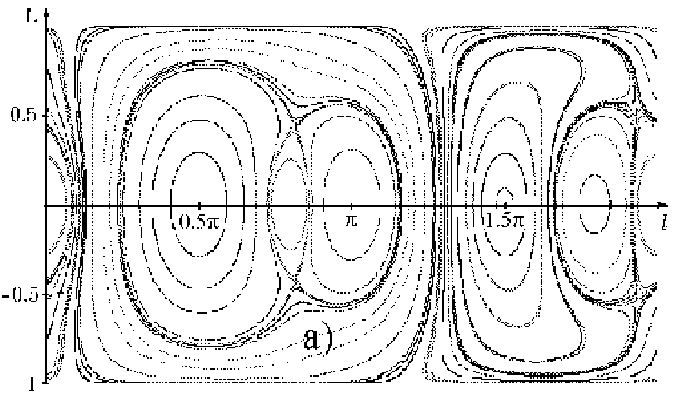}
\includegraphics{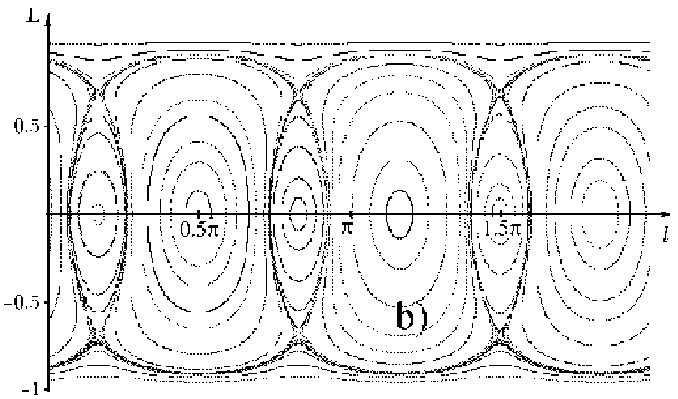}
$$
\caption{The phase portrait of motion of four vortices on a plane with
zero total intensity for
a) $a1\ne a2\ne a3\ne a1$;
b) $a1=a2=a3$.}
\end{figure}

\end{document}